\documentclass[twocolumn]{aastex63}

\usepackage[utf8]{inputenc}
\usepackage{natbib}
\usepackage{graphicx}
\usepackage{soul}
\usepackage{amsmath}
\usepackage{booktabs}
\usepackage{longtable}
\usepackage{xspace}
\usepackage{hyperref}
\usepackage[T1]{fontenc}
\usepackage{lipsum}
\usepackage{comment}
\usepackage{xcolor}
\usepackage{multirow}
\usepackage{IEEEtrantools}
\usepackage{amsmath}

\newcommand{\feh}{\ensuremath{\left[{\rm Fe}/{\rm H}\right]}\,}

\newcommand{\teff}{\ensuremath{T_{\rm eff}}\,}

\newcommand{\logg}{\ensuremath{\log g_*}\,}

\newcommand{\rp}{\ensuremath{\,R_{\rm P}}}
\newcommand{\ecosw}{\ensuremath{e\cos{\omega_{*}}}}
\newcommand{\esinw}{\ensuremath{e\sin{\omega_{*}}}}

\newcommand{\msun}{\ensuremath{\,M_\Sun}}
\newcommand{\rsun}{\ensuremath{\,R_\Sun}}

\newcommand{\tess}{{\it TESS}\,}

\newcommand{\mstar}{\ensuremath{M_{*}}}
\newcommand{\rstar}{\ensuremath{R_{*}}}

\newcommand{\vsini}{\ensuremath{v\sin{i}}\,}
\newcommand{\exofasttwo}{{\tt EXOFASTv2}\,}
\newcommand{\degree}{\ensuremath{\,^{\circ}}}

\newcommand{\obliquity}{\ensuremath{{-0.3}\pm2.2\degree}}

\begin{document}

\title{TOI-1670 c, a 40-day Orbital Period Warm Jupiter in a Compact System, is Well-aligned}

\author[0000-0001-8342-7736]{Jack Lubin}
\affiliation{Department of Physics \& Astronomy, The University of California Irvine, Irvine, CA 92697, USA}

\author[0000-0002-0376-6365]{Xian-Yu Wang}
\affiliation{Department of Astronomy, Indiana University, Bloomington, IN 47405, USA}

\author[0000-0002-7670-670X]{Malena Rice}
\affiliation{Department of Astronomy, Yale University, New Haven, CT 06511, USA}

\author[0000-0002-3610-6953]{Jiayin Dong}
\altaffiliation{Flatiron Research Fellow}
\affiliation{Center for Computational Astrophysics, Flatiron Institute, 162 Fifth Avenue, New York, NY 10010, USA}

\author[0000-0002-7846-6981]{Songhu Wang}
\affiliation{Department of Astronomy, Indiana University, Bloomington, IN 47405, USA}

\author[0000-0002-0015-382X]{Brandon T. Radzom}
\affiliation{Department of Astronomy, Indiana University, Bloomington, IN 47405, USA}

\author[0000-0003-0149-9678]{Paul Robertson}
\affiliation{Department of Physics \& Astronomy, The University of California Irvine, Irvine, CA 92697, USA}

\author[0000-0001-7409-5688]{Gudmundur Stefansson}
\affiliation{Department of Astrophysical Sciences, Princeton University, 4 Ivy Lane, Princeton, NJ 08540, USA}
\affiliation{Henry Norris Russell Fellow}

\author[0000-0003-0353-9741]{Jaime A. Alvarado-Montes}
\affiliation{School of Mathematical and Physical Sciences, Macquarie University, Balaclava Road, North Ryde, NSW 2109, Australia}
\affiliation{The Macquarie University Astrophysics and Space Technologies Research Centre, Macquarie University, Balaclava Road, North Ryde, NSW 2109, Australia}

\author[0000-0001-7708-2364]{Corey Beard}
\altaffiliation{NASA FINESST Fellow}
\affiliation{Department of Physics \& Astronomy, The University of California Irvine, Irvine, CA 92697, USA}

\author[0000-0003-4384-7220]{Chad F.\ Bender}
\affiliation{Steward Observatory, University of Arizona, 933 N.\ Cherry Ave, Tucson, AZ 85721, USA}

\author[0000-0002-5463-9980]{Arvind F.\ Gupta}
\affiliation{Department of Astronomy \& Astrophysics, 525 Davey Laboratory, The Pennsylvania State University, University Park, PA, 16802, USA}
\affiliation{Center for Exoplanets and Habitable Worlds, 525 Davey Laboratory, The Pennsylvania State University, University Park, PA, 16802, USA}

\author[0000-0003-1312-9391]{Samuel Halverson}
\affiliation{Jet Propulsion Laboratory, California Institute of Technology, 4800 Oak Grove Drive, Pasadena, California 91109}

\author[0000-0001-8401-4300]{Shubham Kanodia}
\affiliation{Earth and Planets Laboratory, Carnegie Institution for Science, 5241 Broad Branch Road, NW, Washington, DC 20015, USA}

\author[0000-0001-7318-6318]{Dan Li}
\affiliation{NSF's National Optical-Infrared Astronomy Research Laboratory, 950 N.\ Cherry Ave., Tucson, AZ 85719, USA}

\author[0000-0002-9082-6337]{Andrea S.J.\ Lin}
\affiliation{Department of Astronomy \& Astrophysics, 525 Davey Laboratory, The Pennsylvania State University, University Park, PA, 16802, USA}
\affiliation{Center for Exoplanets and Habitable Worlds, 525 Davey Laboratory, The Pennsylvania State University, University Park, PA, 16802, USA}

\author[0000-0002-9632-9382]{Sarah E.\ Logsdon}
\affiliation{NSF's National Optical-Infrared Astronomy Research Laboratory, 950 N.\ Cherry Ave., Tucson, AZ 85719, USA}

\author[0000-0003-0790-7492]{Emily Lubar}
\affiliation{McDonald Observatory and Department of Astronomy, The University of Texas at Austin, 2515 Speedway, Austin, TX 78712, USA}

\author[0000-0001-9596-7983]{Suvrath Mahadevan}
\affiliation{Department of Astronomy \& Astrophysics, 525 Davey Laboratory, The Pennsylvania State University, University Park, PA, 16802, USA}
\affiliation{Center for Exoplanets and Habitable Worlds, 525 Davey Laboratory, The Pennsylvania State University, University Park, PA, 16802, USA}
\affiliation{ETH Zurich, Institute for Particle Physics \& Astrophysics, Zurich, Switzerland}

\author[0000-0001-8720-5612]{Joe P.\ Ninan}
\affiliation{Department of Astronomy and Astrophysics, Tata Institute of Fundamental Research, Homi Bhabha Road, Colaba, Mumbai 400005, India}

\author[0000-0002-2488-7123]{Jayadev Rajagopal}
\affiliation{NSF's National Optical-Infrared Astronomy Research Laboratory, 950 N.\ Cherry Ave., Tucson, AZ 85719, USA}

\author[0000-0001-8127-5775]{Arpita Roy}
\affiliation{Space Telescope Science Institute, 3700 San Martin Dr, Baltimore, MD 21218, USA}
\affiliation{Department of Physics and Astronomy, Johns Hopkins University, 3400 N Charles St, Baltimore, MD 21218, USA}

\author[0000-0002-4046-987X]{Christian Schwab}
\affiliation{School of Mathematical and Physical Sciences, Macquarie University, Balaclava Road, North Ryde, NSW 2109, Australia}

\author[0000-0001-6160-5888]{Jason T.\ Wright}
\affiliation{Department of Astronomy \& Astrophysics, 525 Davey Laboratory, The Pennsylvania State University, University Park, PA, 16802, USA}
\affiliation{Center for Exoplanets and Habitable Worlds, 525 Davey Laboratory, The Pennsylvania State University, University Park, PA, 16802, USA}
\affiliation{Penn State Extraterrestrial Intelligence Center, 525 Davey Laboratory, The Pennsylvania State University, University Park, PA, 16802, USA}

\begin{abstract}
    We report the measurement of the sky-projected obliquity angle $\lambda$ of the Warm Jovian exoplanet TOI-1670 c via the Rossiter-McLaughlin effect as part of the Stellar Obliquities in Long-period Exoplanet Systems (SOLES) project. We observed the transit window during UT 20 April 2023 for 7 continuous hours with NEID on the 3.5\,m WIYN Telescope at Kitt Peak National Observatory. TOI-1670 hosts a sub-Neptune ($P\sim$11 days; planet b) interior to the Warm Jovian ($P\sim$40 days; planet c), which presents an opportunity to investigate the dynamics of a Warm Jupiter with an inner companion. Additionally, TOI-1670 c is now among the longest-period planets to date to have its sky-projected obliquity angle measured. We find planet c is well-aligned to the host star, with $\lambda = \obliquity$. TOI-1670 c joins a growing census of aligned Warm Jupiters around single stars and aligned planets in multi-planet systems.
\end{abstract}

\section{Introduction}
\par The spin-orbit angle of a planet is a fundamental parameter of the system architecture, as it yields insights into a planet's dynamical evolution and therefore its formation history. While the stellar spin axis is only tilted by $\sim7^{\circ}$ relative to the net orbital angular momentum vector of the solar system \citep{BeckGiles2005}, many exoplanets are found to be misaligned, some severely \citep{Albrecht2022}. The mechanisms that cause misalignment of the star-planet spin-orbit angle are still not fully understood, but many testable hypotheses have been put forth, summarized in \citet{Wang2021}.

\par Most fundamentally, it is unclear whether misalignment is primordial, or a secondary effect born out of planetary dynamics, or if both paths are at play. Two ways to test this are through measuring the spin-orbit angle of Warm Jupiters (defined in this work as $M_p > 0.75\, M_J$ and $a/R_* > 11$) and multi-transiting planet systems. Because of their wide separation, Warm Jupiters cannot impart strong tidal interactions to realign their host stars within the lifetime of the system \citep{Albrecht2012}, even when the host star falls below the Kraft Break \citep[$T_{\mathrm{eff}}\lesssim6100$ K;][]{Kraft1967, Spalding2022}.  On the same note, multi-transiting planet systems likely formed dynamically cold with small planet masses on long orbital periods inhibiting strong interactions. This would make them unlikely to attain spin-orbit misalignment through dynamical disruption and similarly unable to realign their host star in the event of a misalignment. Therefore, Warm Jupiters and/or multiple transiting planet systems very likely retain their primordial spin-orbit angle. In line with this, recently \citet{Rice2022} showed through a compilation of literature measurements that Warm Jupiters orbiting single stars are overwhelmingly aligned. 

\par The Transiting Exoplanet Survey Satellite \citep[TESS;][]{Ricker2015} is delivering many thousands of new multi-planet systems orbiting bright host stars that are amenable to follow up with radial velocity (RV) observations. By extension, the planets in these systems are prime targets for stellar obliquity measurements via the Rossiter-McLaughlin (RM) effect. In this work, which expands upon the investigation of Warm Jupiter obliquities as the ninth installment of the Stellar Obliquities in Long-period Exoplanet Systems program \citep[SOLES;][]{Rice2021, Rice2022, Rice2023, Wang2022, Hixenbaugh2023, Dong2023, Wright2023, Rice2023}, we investigate a system that contains two confirmed transiting exoplanets, one of which is a Warm Jupiter. TOI-1670 is a bright (V=9.9) F7V star \citep{PecautMamajek} that hosts a transiting 11-day orbital period sub-Neptune (b) and a transiting 40-day orbital period Jovian (c) \citep{Tran2022}. Here we present the obliquity measurement via the RM effect for the Jovian planet c. 

\par TOI-1670 c sits in a particularly interesting area of parameter space for obliquity measurements. It orbits its host star at wide enough separation that possible re-alignment through tidal dissipation is highly unlikely. Additionally, the presence of a second transiting planet in the system, the inner sub-Neptune, nearly rules out the possibility of a devastating scattering history that could produce misalignments. The combination of these factors make this obliquity measurement particularly useful to inform the primordial distribution of system alignments in the absence of post-disk dynamical upheavals.

\par This work is laid out as follows. In \S2 we describe the observations. In \S3, we describe the properties of the stellar host. In \S4 we model the RM anomaly and measure the planet's spin-orbit angle. In \S5 we discuss the system in the larger context of Warm Jupiters and multi-planet system obliquity measurements before concluding in \S6. 

\section{Observations}
\label{Observations}

\par We observed a transit of TOI-1670 c on UT 2023 April 20 with NEID \citep{Schwab2016} on the WIYN\footnote{The WIYN Observatory is a joint facility of the NSF's National Optical-Infrared Astronomy Research Laboratory, Indiana University, the University of Wisconsin-Madison, Pennsylvania State University, and Purdue University.} 3.5 m telescope at Kitt Peak National Observatory. NEID is a fiber-fed \citep{kanodia2018, Kanodia2023}, ultra-stable \citep{Stefannson2016, Robertson2019} spectrograph spanning the 380-930 nm wavelength range with a resolving power of $R\sim$110,000 in its high resolution mode \citep{Halverson2016}. With an exposure time of 1200 seconds, we obtained 21 spectra over a $\sim$7 hour span of observations, beginning at airmass = 1.99 and continually rising until observations end at airmass = 1.32. Fifteen of these spectra were obtained within the 5 hour transit window between 06:18 UT and 11:29 UT, while the other six were obtained prior to the start of the transit. 

\par Because $12^{\circ}$ twilight occurred within 20 minutes after the egress, we opted to obtain 2 hours of out-of-transit baseline prior to ingress. Queue observers reported that conditions at the start of observations were not ideal, with transient cloud cover passing over. However, directly before ingress, conditions improved and the remainder of the night was clear with stable seeing of $\sim$1\arcsec{}. This can be seen in the improvement of the uncertainty estimates after the fourth observation, see Figure \ref{fig:fits}. NEID's standard afternoon and morning calibration sequences were obtained at the beginning and end of the night. Additionally, we obtained 7 Fabry-P\'erot etalon exposures during our observations for instrumental drift correction. The first etalon exposure was taken immediately prior to our first target exposure, and 6 additional etalon exposures were interspersed with throughout our observations, evenly spaced out at roughly hour intervals. 

\par To extract the RVs from the spectra, we used a modified version of the \texttt{SERVAL} code \citep{Zechmeister2018} following the procedures described in detail within \citet{Stefansson2022}. A comparison to the RVs derived from the standard NEID Data Reduction Pipeline, computed via the Cross-Correlation Function (CCF) method \citep{Baranne1996, Pepe2002}, showed that all observations except the very last one were in agreement to within $1\sigma$. The last observation was taken right up against morning $12^{\circ}$ twilight, and its value is $>5\sigma$ discrepant between the CCF to the \texttt{SERVAL} reductions. 

\par To attempt to soften this discrepancy, we re-reduced the data with \texttt{SERVAL} and excluded the bluest orders, using only information from wavelengths 4260\r{A} - 8940\r{A} to calculate the RV. Across all observations, the excluded bluest orders all had a signal-to-noise ratio (S/N) $<$ 13, while the included orders have an average S/N of 31. This was in an effort to mitigate any potential stray light from twilight. Ultimately, this effort did not prove to re-align the last point (still  $>5\sigma$ discrepant). Therefore, we exclude this last data point from our analysis. Ultimately, we chose to analyze a \texttt{SERVAL} reduction over a CCF reduction due to higher precision RVs (median uncertainty values of 4.8 m/s vs. 6.4 m/s), and we chose to analyze the blue-excluded \texttt{SERVAL} reduction over the standard because the bluest orders are low S/N such that they contribute red noise to the time series.

\section{Stellar parameters}

\subsection{Synthetic spectral fitting by \texttt{iSpec}}

\par The 20 spectroscopic observations of TOI-1670 from NEID that were used to model the RM anomaly also enabled us to determine the star's spectroscopic parameters, including stellar effective temperature  (\teff), surface gravity (\logg), metallicity (\feh), and projected rotational velocity (\vsini). We used the synthetic spectral fitting technique provided by the Python package  \texttt{iSpec} \citep{Blanco2014,Blanco2019} to measure these parameters.

\par Specifically, we used the SPECTRUM radiative transfer code \citep{Gray1994}, the MARCS atmosphere model \citep{gustafsson2008_MARCS}, and the sixth GES atomic line list \citep{Heiter2021_GES} integrated within \texttt{iSpec}, to generate a synthetic model for co-added NEID spectra with a S/N of 201. Micro-turbulent velocities were considered as a variable parameter in the fitting procedure, providing the adaptability to represent the small-scale turbulent motions within the stellar atmosphere. On the other hand, macro-turbulent velocities were ascertained through an empirical relationship \citep{Doyle2014Vmac}, utilizing well-established correlations with various stellar characteristics. Special regions have been chosen to expedite the fitting progress. These regions encompass the wing segments of the H$\alpha$, H$\beta$, and Mg I triplet lines, which are sensitive to $T_{\mathrm{eff}}$ and $\log g$, and also include the Fe I and Fe II lines, which provide precise constraints on [Fe/H] and $v\sin i$. We then employed the Levenberg-Marquardt nonlinear least-squares fitting algorithm \citep{Markwardt2009} to iteratively minimize the $\chi^2$  value between the synthetic and observed spectra. The resulting spectroscopic parameters are listed in Table~\ref{tab:results}. 

\subsection{SED+MIST fit by \texttt{EXOFASTv2}}

To determine additional stellar parameters, including the stellar mass (\mstar) and radius (\rstar), we employed the MESA Isochrones \& Stellar Tracks model \citep{Choi2016mist,Dotter2016mist}, coupled with a spectral energy distribution (SED) fit. We assembled broad photometry using multiple catalogs, which include 2MASS \citep{Cutri2003}, WISE \citep{Cutri2014AllWISE}, TESS \citep{Ricker2015}, and Gaia DR2 \citep{GaiaCollaboration2018}. We adopted Gaussian priors on \teff and \feh, derived from our synthetic spectral fitting, as well as parallax from Gaia DR3 \citep{GaiaCollaboration2023} and $V$-band extinction from TIC 8.2 catalog \citep{Paegert2021arXiv}. Note that we inflated the uncertainty of \teff \, to 150 K to account for the roughly 2.4\% systematic uncertainty floor for \teff \, indicated by \cite{Tayar2020}. For the SED fit, we used the Differential Evolution Markov Chain Monte Carlo (DEMCMC) method integrated into \exofasttwo \citep{Eastman2017,Eastman2019} to estimate the uncertainties. We deemed the MCMC process converged when the Gelman-Rubin diagnostic \citep[$\hat{R}$;][]{Gelman1992} was less than 1.01 and the number of independent draws exceeded 1000. Our final stellar parameters as well as the values from the discovery paper \citep{Tran2022} for comparison are listed in Table~\ref{tab:results}. The agreements for stellar parameters from our work and discovery paper \citep{Tran2022} are less than 1 $\sigma$ except \teff ($1.75\sigma$) and \feh ($1.5\sigma$). The slight  disagreements on \teff and \feh might be caused by the difference in data quality and spectra analysis software \citep{Blanco2019}.

\label{st_pars}
\begin{deluxetable*}{lllllll}
\tabletypesize{\scriptsize}
\tablecaption{Priors and posteriors for the TOI-1670 planetary system.}\label{tab:results}
\tablehead{
\colhead{ } & \colhead{Description (units)} &\colhead{Priors$^{a}$} &\colhead{Fitted Value}&\colhead{ } & \colhead{ \cite{Tran2022}}}
\tablewidth{300pt}
\startdata
\multicolumn{5}{l}{\textbf{Stellar Parameters:}}\\
 &      &      &   Spectrum fit   & SED+MIST \\
  &      &      &      &   (adopted)\\
\,\,$M_*$  & Mass (\msun)  & - &-& $1.219^{+0.059}_{-0.070}$ & $1.21\pm0.02$\\ \
$R_*$  & Radius (\rsun)  & - &-& $1.308^{+0.031}_{-0.029}$& $1.316\pm0.019$\\ \
$\log{g_*}$  & Surface gravity (cgs)  & -&$4.26\pm0.20$& $4.290^{+0.030}_{-0.034}$& $4.29\pm0.11$\\ \
$T_{\rm eff}$  & Effective Temperature (K)  & $\mathcal G(6204;150)$ &$6328\pm96$& $6330^{+68}_{-70}$& $6170\pm61$\\ \
$[{\rm Fe/H}]$  & Metallicity (dex)  & $\mathcal G(-0.028;0.036)$ &$-0.01\pm0.07$& $0.017^{+0.054}_{-0.048}$& $0.09\pm0.007$\\ \
$v \sin{i_*}$& Host star projected rotational velocity (km/s)      & -   &     $8.13\pm1.02$      &-   &- \\
$Age$  & Age (Gyr)  & - &-&$2.3^{+1.7}_{-1.1}$& $2.53\pm0.43$\\ \
$A_V$  & V-band extinction (mag)  & $\mathcal G(0.038;0.020)$ &-& $0.075^{+0.031}_{-0.042}$   &-\\ \
$\varpi$  & Parallax (mas)  & $\mathcal G(6.022;0.013)$ &-& $6.022\pm0.013$   &-\\ \
$d$  & Distance (pc)  & - &-& $166.04\pm0.34$ & -\\ \
\\
\hline \\
\multicolumn{5}{l}{\textbf{Rossiter-McLaughlin Parameters:}}\\
 &      &      &   Allesfitter   & rmfit\\
  &      &      &   (adopted)   & \\
$\lambda$& Sky-projected spin-orbit angle (deg)      & $\mathcal U(0;0;180)$   &   $-0.3\pm2.2$        & $0.2\pm2.6$&-\\
$v \sin{i_*}$& Host star projected rotational velocity (km/s)      & $\mathcal U(9.2;0;20)$   &     $8.95\pm0.46$      & $9.54_{-1.00}^{+1.00}$&$9.2\pm0.6$ \\
$\beta$ & Intrinsic stellar line width (km/s) & $\mathcal G(6.0;1.0)$ &  - & $6.0\pm1.0$&-\\
$\zeta$ & Macro-turbulent velovity (km/s) & $\mathcal G(1.32;1.0)$ &  $1.31_{-0.78}^{+0.94}$ & -&-\\
$\xi$ & Micro-turbulent velovity (km/s) & $\mathcal G(5.41;1.0)$ &  $5.35\pm0.92$ & -&-\\
\hfill \\
\hline \\
\multicolumn{5}{l}{\textbf{Planetary Parameters:}}\\
$R_c / R_\star$&     Planet-to-star radius ratio&  $\mathcal G(0.077;0.002)$   &   $0.07616\pm0.00044$   & $0.0769_{-0.0020}^{+0.0020}$ &$0.077\pm0.002$\\
$(R_\star + R_c) / a_c$&     Sum of radii divided by orbital semi-major axis&   $\mathcal G(0.02647;0.00004)$   &   $0.026470\pm0.000040$     &$0.02647_{-0.00004}^{+0.00004}$ &-\\
$\cos{i_c}$&   Cosine of the orbital inclination&    $\mathcal G(0.0204;0.0007)$   &   $0.02026\pm0.00015$     & $0.02153_{-0.00058}^{+0.00069}$&-\\
$P_c$& Orbital period (days)      & $\mathcal G(40.74976;0.0002)$   &     $40.750145\pm0.000025$     & $40.750198_{-0.00006}^{+0.00006}$ &$40.74976_{-0.00021}^{+0.000022}$&-\\
$T_{0;c}$& Transit epoch - 2459000(BJD)      & $\mathcal G(402.87902;0.01)$   &     $402.88333\pm0.00024$      & $402.88389_{-0.00034}^{+0.00034}$ &-\\
$K_c$& Radial velocity semi-amplitude (km/s)      & $\mathcal G(32.7;4.7)$   &   $30.1\pm2.4$    & $32.7\pm4.7$& $32.7_{-4.3}^{+4.7}$\\
$\sqrt{e_c} \cos{\omega_c}$&  & $\mathcal G(-0.08;0.15)$   &      $-0.076\pm0.081$     & $-0.1_{-0.16}^{+0.16}$& $-0.07_{-0.13}^{+0.14}$\\
$\sqrt{e_c} \sin{\omega_c}$&  & $\mathcal G(0.29;0.09)$   &     $0.235_{-0.039}^{+0.034}$     & $0.314_{-0.093}^{+0.096}$& $0.27_{-0.1}^{+0.08}$\\
$q_{1;\rm NEID}$& Linear limb-darkening coefficient for NEID&  $\mathcal G(0.32;0.1)$   &  $0.353\pm0.093$    & $0.327_{-0.097}^{+0.098}$ &-\\
$q_{2;\rm NEID}$& Quadratic limb-darkening coefficient for NEID&  $\mathcal G(0.22;0.1)$   &  $0.301\pm0.089$    & $0.232_{-0.100}^{+0.098}$ &-\\
$q_{1;\rm TESS}$& Linear limb-darkening coefficient for TESS&  $\mathcal G(0.30;0.1)$   &   $0.235_{-0.040}^{+0.043}$    & -&$0.35_{-0.11}^{+0.19}$\\
$q_{2;\rm TESS}$& Quadratic limb-darkening coefficient for TESS&  $\mathcal G(0.22;0.1)$   &   $0.221\pm0.094$   & -&$0.32_{-0.23}^{+0.39}$\\
$\ln{\sigma_\mathrm{TESS}}$ &Jitter term for TESS ($\ln{ \mathrm{km/s}}$)& $\mathcal U(-3;-15;0)$& $-7.956\pm0.035$ &-&-  \\
$\ln{\sigma_\mathrm{jitter; NEID}}$  &Jitter term for NEID ($\ln{ \mathrm{km/s}}$)&$\mathcal U(-3;-15;0)$&$-10.4\pm3.3$ & -&- \\
$\ln{\sigma_\mathrm{jitter; HARPS-N}}$  &Jitter term for HARPS-N ($\ln{ \mathrm{km/s}}$)&$\mathcal U(-3;-15;0)$& $-10.0\pm3.4$ & -&- \\
$\ln{\sigma_\mathrm{jitter; FIES}}$  &Jitter term for FIES ($\ln{ \mathrm{km/s} }$)&$\mathcal U(-3;-15;0)$& $-10.1\pm3.4$ & -&- \\
$\ln{\sigma_\mathrm{jitter; TULL}}$  &Jitter term for TULL ($\ln{ \mathrm{km/s} }$)&$\mathcal U(-3;-15;0)$& $-3.82\pm0.16$ &- &- \\
\\
\hline \\
\multicolumn{5}{l}{\textbf{Derived Parameters}:}\\
$M_c$& Planetary mass (M$_\mathrm{jup}$)&  -   &   $0.578_{-0.055}^{+0.059}$   & - & $0.63_{-0.08}^{+0.09}$\\
$R_c$& Planetary radius (R$_\mathrm{jup}$)&  -   &    $0.970\pm0.023$    & -& $0.987_{-0.025}^{+0.025}$\\
$a_c / R_\star$& Semi-major axis over host radius&  -   &   $40.656\pm0.064$    & -& $40.68_{-0.66}^{+0.66}$\\
$i_c$& Inclination&  -   &   $88.8390\pm0.0088$    & -& $88.84_{-0.04}^{+0.04}$\\
$e_c$& Eccentricity&  -   &   $0.067_{-0.018}^{+0.019}$    & -& $0.09_{-0.04}^{+0.05}$\\
$\omega_b$& Argument of periastron&  -   &   $108\pm19$    & -&-\\
$T_\mathrm{tot;c}$& Total transit duration (hours)&  -   &   $5.392\pm0.029$    & -&-\\
$b$& Impact parameter&  -   &   $0.7733_{-0.010}^{+0.0092}$    & -& $0.76_{-0.04}^{+0.02}$\\
$u_\mathrm{1; NEID}$ & Linear limb-darkening coefficient 1 for NEID&  -   &   $0.35_{-0.11}^{+0.12}$    & - &-\\
$u_\mathrm{2; NEID}$ & Quadratic limb-darkening coefficient 2 for NEID&  -   &    $0.23_{-0.10}^{+0.11}$   & -&-\\
$u_\mathrm{1; TESS}$ & Linear limb-darkening coefficient 1 for TESS&  -   &    $0.213\pm0.090$    & -&-\\
$u_\mathrm{2; TESS}$ & Quadratic limb-darkening coefficient 2 for TESS&  -   &   $0.269_{-0.097}^{+0.10}$   & -&-\\
\enddata
\tablenotetext{a}{For stellar parameters, the priors were only applied for the SED+MIST fit.}
\end{deluxetable*}

\section{Stellar obliquity derivation} 

Visual inspection of the NEID RVs reveals a clear, symmetric RV anomaly during transit, indicative of a well-aligned orbit. To measure the sky-projected spin-orbit angle ($\lambda$), we employed a modified version of \texttt{allesfitter} \citep{Max2021} to perform a global fit of TOI-1670 c, which included an RM measurement from the NEID spectrograph, 14 TESS\footnote{DOI: 10.17909/amqq-ke07} transits from Sectors 16, 18, 19, 21, 23, 25, 40, 41, 47, 49, 50, 52, 58, and 59, and out-of-transit RVs from the FIES, HARPS-N, and Tull spectrographs as presented by \cite{Tran2022}. The 2-minute cadence Pre-search Data Conditioning Simple Aperture Photometry \cite[PDCSAP;][]{Smith2012PDCSAP,Stumpe2012, Stumpe2014} light curve was adopted in our work, which is generated by the Science Processing Operation Center \cite[SPOC;][]{Jenkins2016} team. We downloaded it via \texttt{lightkurve} \citep{Lightkurve2018} package and excluded data points with severe quality issues.

In \texttt{allesfitter}, the model for transit, radial velocity, and Rossiter-McLaughlin effect is implemented via \texttt{ellc} \citep{ellc}. The RM effect model provided by \texttt{ellc} is calculated using the flux-weighted sum of the radial velocities over the visible part of the star, which does not account for instrumental broadening, macroturbulence, and other broadening factors. To address this limitation, we replaced the \texttt{ellc} RM model with the analytic model presented by \citet{Hirano2011}, as implemented in \texttt{\st{tracit}} \citep{Hjorth2021,KnudstrupAlbrecht2022}.

In this fit, we assumed Gaussian priors for various parameters: the orbital period ($P$), the reference transit mid-time ($T_0$), the cosine of the orbital inclination ($\cos i$), the planet-to-star radius ratio ($\rp / \rstar$), the sum of radii divided by the orbital semi-major axis ($(\rstar + \rp )/ a$), the RV semi-amplitude ($K$), and the parameterized eccentricity and argument of periastron ($\ecosw$ and $\esinw$). These priors were taken from the discovery paper \citep{Tran2022}. We also included uniform priors for jitter terms (ln$ \sigma_{\rm jitter}$) for RV datasets and error scaling factors (ln$ \sigma$) for \tess photometry. 

For the sky-projected stellar rotational velocity (\vsini) and the spin-orbit angle ($\lambda$), we adopted uniform priors: $\mathcal{U}(0;20)$ for \vsini and $\mathcal{U}(-180;+180)$ for $\lambda$. The initial estimates for these parameters were set at 9.2 km/s and $0^{\circ}$, respectively. Note that $T_0$ was repositioned to the middle of the temporal baseline to reduce the degeneracy between the orbital period and epoch. 

Additionally, we adopted priors on macro-turbulence ($\zeta$) and micro-turbulence ($\xi$) to account for stellar surface motion. The priors on these parameters were derived from \cite{Doyle2014Vmac} and \cite{Jofre2014Vmic}, respectively, and a standard deviation ($\sigma$) of 1 km/s was employed. Moreover, transformed limb darkening coefficients for \emph{TESS} ($q_{\rm 1:TESS}$, $q_{\rm 2:TESS}$) and NEID ($q_{\rm 1:NEID}$, $q_{\rm 2:NEID}$) were considered. With the stellar parameters determined by the SED+MIST fit, limb-darkening parameters for these two instruments were estimated via the \texttt{QUADLD} function embedded in \texttt{EXOFASTv2}. For NEID, the limb darkening coefficients were computed as the mean of the coefficients from the $R$ and $I$ bands, which encompass the majority of the RV information content for these data. 

We used a third-order polynomial function for each transit to model potential trends. A constant baseline and a jitter term were included to account for the RV offset in the out-of-transit RVs. Specifically, a quadratic function was used for the RM fit to model short-term overnight instrumental systematics and stellar variability. To account for the 1200s exposure time of our RM observations, exposure interpolation was performed during the fit. The number of fine sampling points was set as 10. To estimate the posterior distributions of system parameters, we applied the Affine-Invariant Markov Chain Monte Carlo \citep{Goodman2010} algorithm embedded in \texttt{emcee} \citep{Dan2013}. The chain ran with 100 walkers for 200,000 steps, and the resulting number of independent draws exceeded 100, marking convergence. The system parameters derived from this fit are presented in Table~\ref{tab:results}, most notably the sky-projected spin-orbit angle is measured to be \obliquity. 

\begin{figure*}[t]
    \centering\includegraphics[width=0.91\textwidth]{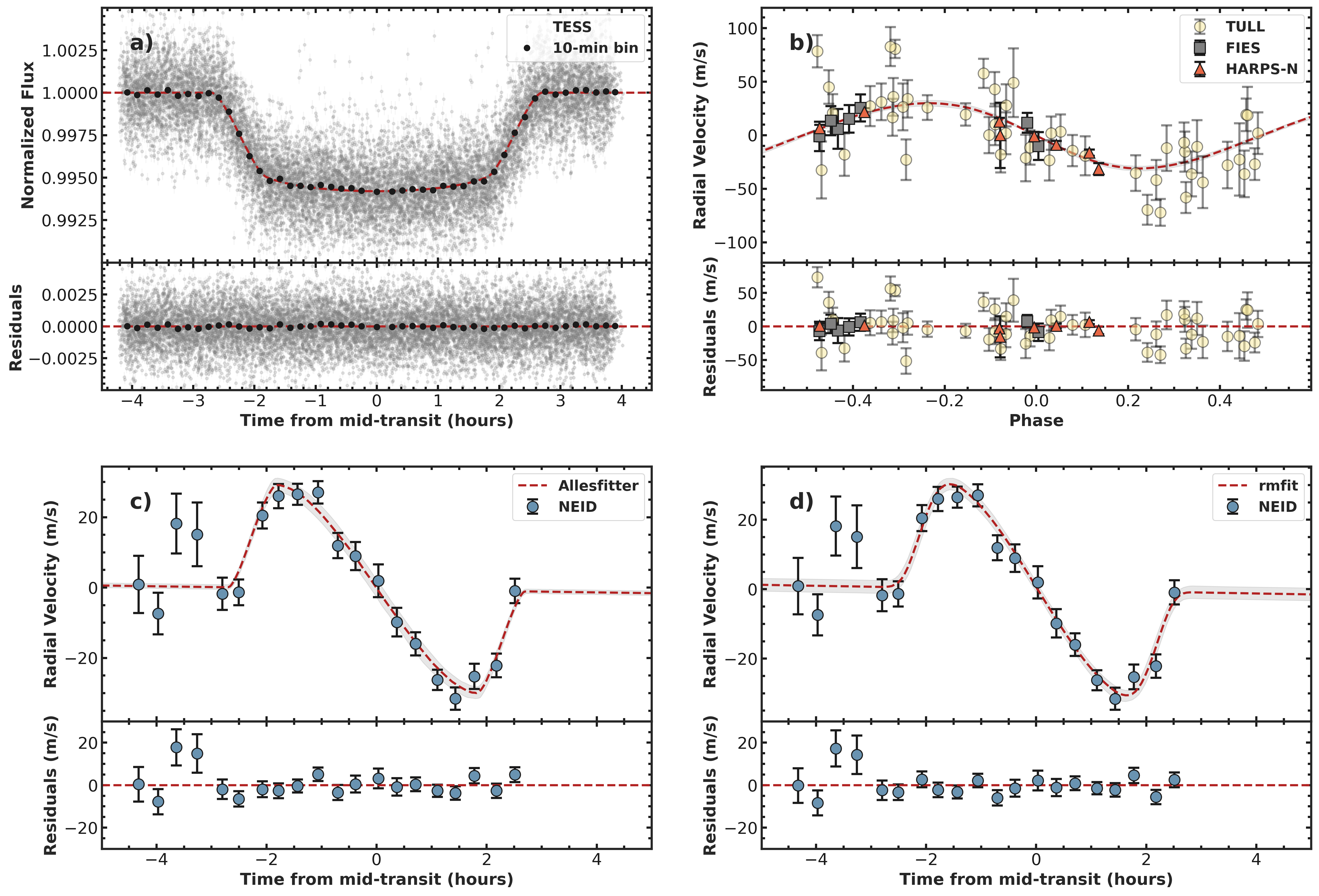}
    \caption{A combined fit of photometry, non-transit radial velocity data, and Rossiter-McLaughlin measurements was conducted for TOI-1670 c. Panel \textit{a} displays the modeling of $\tess$ transits. The grey dots represent the phase-folded transit, while the red dashed line illustrates the modeled light curve. Panel \textit{b} presents the radial velocity fit using RV data from various spectrographs, namely TULL, FIES, and HARPS-N. The observed data is differentiated by unique markers and colors for each instrument, and the red dashed line depicts the modeled RV curve. Panels \textit{c} and \textit{d} detail the RM modeling based on \texttt{allesfitter} and \texttt{rmfit}, respectively. The RM data are represented by blue dots, and the RM model is shown in red. Grey regions indicate the 1$\sigma$ credible intervals. Residuals are displayed beneath each panel.}
    \label{fig:fits}
\end{figure*}

We also conducted an alternative RM analysis using the \texttt{rmfit} code \citep{Stefansson2020rmfit}, which incorporates the RM model from \cite{Hirano2011} and the RV model from \texttt{radvel} code \citep{Fulton2018}. The same priors and supersampling strategy that were used in the \texttt{allesfiter} fit were adopted. Unlike the use of $\zeta$ and $\xi$ in the \texttt{allesfiter} fit, we designated the intrinsic line width, $\beta$ (which is related to micro-turbulence, see \citealt{Hirano2011}), to match the width of the NEID resolution element \citep{Boue2013}: specifically, $\beta = 6.0 \pm 1.0$ km/s. This integrated uncertainty is intended to consider possible impacts from macro-turbulence or other factors that could widen the line profile. Firstly, we employed the \texttt{PyDE} differential evolution optimizer \citep{pyde} to identify a global maximum-likelihood solution. Then we initialized a set of 100 walkers in the vicinity of this solution and sampled the posteriors with 50,000 steps using \texttt{emcee}. The resulting Gelman-Rubin statistic factors ($\hat{R}$) for all parameters are less than 1.01, and the resulting number of independent draws exceeded 100, which is considered as converged. The resulting parameters are summarized in Table~\ref{tab:results}, with the solution corresponding to a sky-projected spin-orbit angle of $0.2\pm2.6\degree$. See Figure \ref{fig:fits} for the global transit/RV/\texttt{allesfitter} RM model as well as the \citep{Hirano2010} \texttt{rmfit} RM model. Additionally, all planetary parameters derived in our work agree within $1\sigma$ of those from the discovery paper \citep{Tran2022}, with the exception of the orbital period which agrees within $1.82\sigma$. This result is anticipated, as our work incorporated 14 transits, in contrast to the 6 utilized in the discovery paper, thereby yielding a more precise determination of the orbital period. Furthermore, we do not find evidence for any transit timing variations (TTVs) across the $\sim$1200 day \tess\, baseline. Note that the uncertainty associated with $q_{1}$ from NEID RM measurements is double that of \tess\, transits, attributable to the limited sample rate and precision of these measurements. However, the transformed limb darkening coefficients, $q_{1}$ and $q_{2}$, from NEID RM and \tess\, transits agree very well with each other. They are also consistent with their predicted values based on the stellar parameters.

Given the extensive coverage of the \tess \, light curve (from July 18th, 2019 to January 18th, 2023), we searched for periodic modulations that may be attributable to stellar activity tracing the host star's rotation period. Utilizing \texttt{SpinSpotter} \citep {Holcomb2022}, we were not able to confidently identify a rotation period despite the high stellar $v\sin i$, and therefore we are not able to place limits on the true obliquity of the system.

\section{Discussion}
\label{Discussion}

\par
\begin{figure*}[t]
    \centering\includegraphics[width=0.95\textwidth]{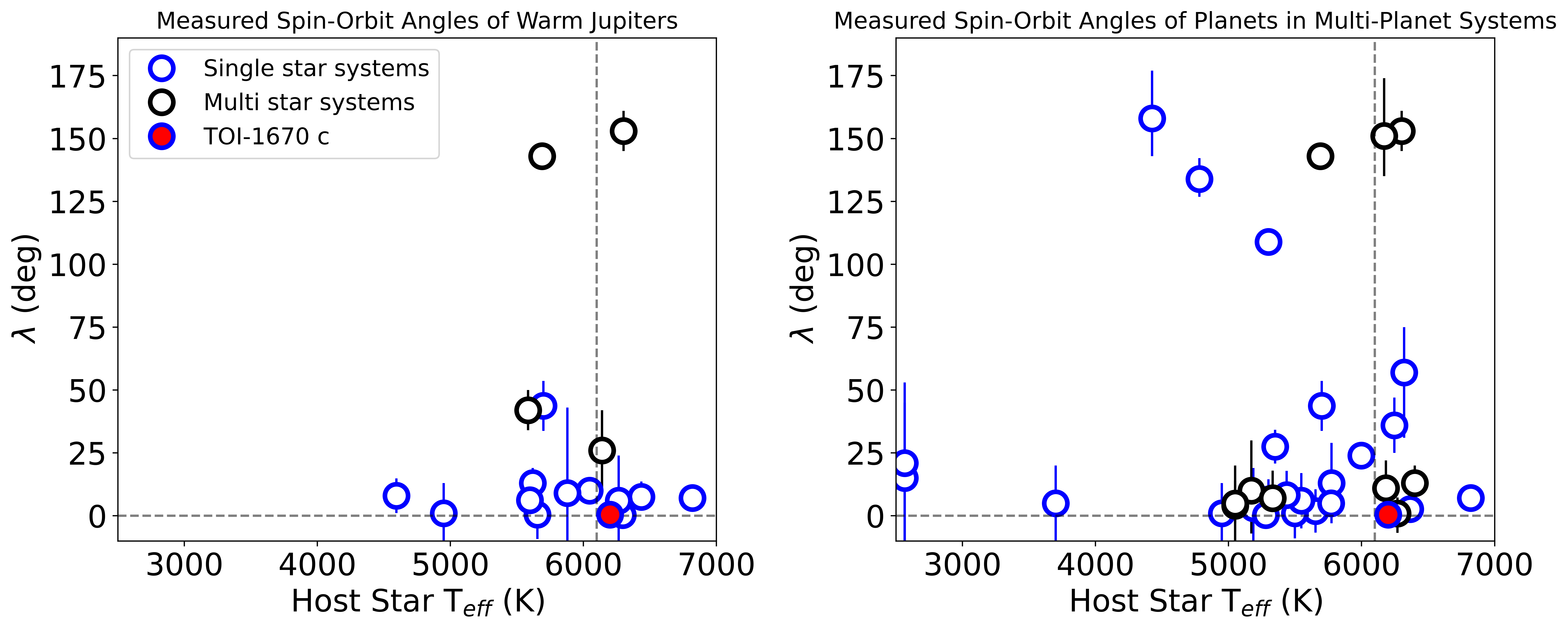}
    \caption{Spin-orbit angles for populations of Warm Jupiters ($M_p > 0.75 M_J$ and $a/R_* > 11$) and more generally for multi-planet systems. Both figures include only measurements with uncertainties less than 40\degree. \textbf{Left:} the population of Warm Jupiters with measured spin-orbit angles, split by stellar host multiplicity. Those around single stars are overwhelmingly aligned, regardless of their temperature in relation to the Kraft break. \textbf{Right:} the population of planets within multi-planet systems (including non-transiting planets) with measured spin-orbit angles. This population is also overwhelmingly well-aligned.}
    \label{fig:angles}
\end{figure*}

 Our result adds to the growing census of Warm Jupiters in single star systems with aligned spin-orbit angles. TOI-1670 c now joins 15 other aligned Warm Jupiters around single star systems, 14 are outlined in \citet{Rice2022}, plus WASP-106 \citep{Harre2023, Wright2023}. Of these 16 systems, only WASP-38, KOI-12, TOI-677 \citep{Sedaghati2023}, and TOI-1670 are above the Kraft break, see Figure \ref{fig:angles}. This is notable because to probe spin-orbit angles of host stars above the Kraft break is to probe a primordial arrangement of the system at the time of formation. Due to their convective envelopes, stars below the Kraft temperature break \citep{Kraft1967, Spalding2022} can realign themselves to a previously misaligned planet through a tidal dissipation mechanism \citep{Albrecht2022}. However, stars above the temperature break cannot realign themselves within the timescale of the system lifetime. In TOI-1670, regardless of the Kraft break, planet c is widely enough separated from the host star that tidal forces cannot play a significant role in re-aligning the planet. For planet c to realign the host star, following Eq 15. from \citet{Winn2005} we compute a realignment timescale of $\sim10^{22} \times Q_*$ years, far older than the age of the universe. Even accounting for uncertainty in the physical parameters and rotation period, allowing for order of magnitude changes of each still constrains the realignment timescale to be much longer than the age of the universe. Therefore, TOI-1670 c's aligned obliquity angle falls neatly into the findings outlined in \citet{Rice2022}, where Warm Jupiters form quiescently in an aligned proto-planetary disk. 

\par \citet{Schlaufman2010} first proposed that the Kraft temperature break actually represents a stellar mass break. Recently, \citet{Hixenbaugh2023} described that the stellar mass break may reflect a break in planet-planet interaction history rather than one of tidal history. Hotter stars (on the Main Sequence) are more massive, and more massive stars form out of more massive disks \citep{Andrews2013} which are more capable of producing multiple massive planets \citep{Yang2020}. With multiple massive planets, gravitational interactions like those described in \citet{FordRasio1996}, \citet{Wu2011}, \citet{Petrovich2015}, and \citet{Naoz2016} are more likely, and such interactions may explain the observed misaligned orbits of Jovian planets. Meanwhile, cooler stars are less massive: they form out of less massive disks where there is not enough material to produce multiple massive planets. Scattering is then less likely and the lone, quiescently formed Jovian retains its primordial spin-orbit angle. This angle need not necessarily be aligned \citep{Watson2011,Hurt2023}, and Lidov-Kozai interactions with a third body can also misalign the primordial system \citep{Naoz2011}. This is further in agreement with observational studies that find Warm Jupiters have high companion rates, but these companions are almost always small planets, not other Jovians \citep{Huang2016, Wu2023}

\par Adding to the intrigue of TOI-1670's aligned stellar obliquity measurement is that it is a part of a compact multi-planet system. In general, very few planets in multi-exoplanet systems have had their sky-projected stellar obliquities measured, now only 33 of 198 planets, see catalog from \citet{Southworth2011} which is continually updated online. Of these 33, only 9 planets in 4 systems have had multiple planet spin-orbit angles from the same system measured via the RM effect: TRAPPIST-1 \citep[all aligned;][]{Hirano2020}, K2-290 \citep[both misaligned;][]{Hjorth2021}, HD 3167 \citep[perpendicular orbits;][]{Bourrier2021}, and V1298 Tau \citep[both aligned;][]{Feinstein2021, Johnson2022}. For three distinct outcomes in the four instances of this measurement, there is a growing need to add to the census of multiply-measured obliquity systems. Although this is a very small sample size, as 2 of 4 host at least one misaligned planet, this group appears to be distinct from the 24 other planets in multi-planet systems where only one planet within the system has had a spin-orbit angle measured: only 6 of 24 are misaligned, $\pi$ Men b \citep{Kunovac2021} and K2-93 d \citep{Grouffal2022}, WASP-134 \citep{Anderson2018}, HAT-P-11 \citep{SanchisOjeda2011}, WASP-8 \citep{Queloz2010}, and WASP-107 \citep{Rubenzahl2021}. For the latter 3, the second planet in the system is a long-period giant planet which may have caused the misalignment. 

\par Given that Warm Jupiters in single-star systems have all been aligned to date \citep{Rice2022}, and planets in multi-transiting systems are nearly all aligned \citep{Albrecht2013, Wang2018, Wang2022}, the architecture of TOI-1670's planetary system would suggest at the surface that the system should be aligned. This is part due to the second transiting planet in the system. \citet{Tran2022} performed the initial characterization, reporting an 11-day sub-Neptune with a mass of $13^{+9.5}_{-8.7} \, M_{\oplus}$. The transiting nature of planets b and c in this system hint at a co-planar architecture, where all planets formed quiescently out of the aligned proto-planetary disk and are therefore themselves individually well-aligned to the stellar spin axis. This picture is fully in agreement with the description in \citet{Wu2023} and \citet{Hixenbaugh2023} where systems with a single Jovian are less likely to have dynamically violent history, and therefore more likely to have all planets well aligned. Additional measurement of the spin-orbit angle for planet b will be required to confirm this architecture.

\par While the current sample size is still small, measurements of stellar obliquity in multi-planet systems so far suggest a trend towards alignment. This is a parameter space that needs to be more fully explored. As Hot Jupiters have dominated the obliquity measurements for the past two decades, \citep[152 of 198 measurements;][]{Southworth2011}, and as Hot Jupiters are found to preferentially occur as single planet systems \citep{Wright2012, Steffen2012, Huang2016, Petigura2018, Wu2023}, the census of measured stellar obliquities is biased against multi-planet systems. In the new era of Extreme Precision Radial Velocity (EPRV) instruments ($<1$ m/s single measurement precision) on large aperture ($>4$ m) telescopes, the small RM signals of sub-Neptunes, often found in compact multi-planet systems \citep{Weiss2018} are now within reach. 

\section{Conclusion}
\label{Conclusion}

\par We observed the RM anomaly for TOI-1670 c, a Warm Jupiter at wide separation within a multi-planet system around a single star. We measure the sky-projected obliquity angle to be $\lambda = \obliquity$. The aligned obliquity is consistent with expectations for a wide-separation planet that formed in a well-aligned proto-planetary disk and stayed aligned.  

\par This measurement is the only Warm Jupiter with a sub-Neptune inner companion to have its spin-orbit angle measured. More samples from these understudied bins of parameter space will be needed to probe the dynamics of these complex, yet ubiquitous systems. 

\section{Acknowledgements}
\label{Acknowledgements}

\par The authors are honored to be permitted to conduct astronomical research on Iolkam Du\'ag (Kitt Peak), a mountain with particular significance to the Tohono O\'odham. We thank the queue observer, Yatrik Patel, and the telescope operators, John Della Costa and Amy Robertson, for their efforts in collecting this data set on our behalf. We also thank the maintenance staff at the WIYN telescope at Kitt Peak National Observatory for their efforts in keeping the telescope in good working condition. We also thank the anonymous referee for their time and constructive feedback. 
\par This work is based on observations at Kitt Peak National Observatory, NSF’s NOIRLab (Prop.~ID 2023A-273787; PI: J. Lubin), managed by the Association of Universities for Research in Astronomy (AURA) under a cooperative agreement with the National Science Foundation. Data presented herein were obtained at the WIYN Observatory from telescope time allocated to NN-EXPLORE through the scientific partnership of the National Aeronautics and Space Administration, the National Science Foundation, and the National Optical Astronomy Observatory. M.R. and S.W. thank the Heising-Simons Foundation for their generous support. M.R. acknowledges support from Heising-Simons Foundation Grant \#2023-4478, as well as the 51 Pegasi b Fellowship Program. 
S.W. acknowledges support from Heising-Simons Foundation Grant \#2023-4050. GS acknowledges support provided by NASA through the NASA Hubble Fellowship grant HST-HF2-51519.001-A awarded by the Space Telescope Science Institute, which is operated by the Association of Universities for Research in Astronomy, Inc., for NASA, under contract NAS5-26555. Xian-Yu thanks to the computational resources provided by Indiana University. This research was supported in part by Lilly Endowment, Inc., through its support for the Indiana University Pervasive Technology Institute. The Center for Exoplanets and Habitable Worlds is supported by Penn State and the Eberly College of Science. This work was performed for the Jet Propulsion Laboratory, California Institute of Technology, sponsored by the United States Government under the Prime Contract 80NM0018D0004 between Caltech and NASA.

\bibliography{bib.bib}

\end{document}